\documentclass[english,aps,twocolumn]{revtex4}
\usepackage[T1]{fontenc}
\usepackage[latin9]{inputenc}
\usepackage{babel}
\usepackage{float}
\usepackage{amsthm}
\usepackage{amsmath}
\usepackage{amssymb}
\usepackage{graphicx}
\usepackage[unicode=true,pdfusetitle,
 bookmarks=true,bookmarksnumbered=false,bookmarksopen=false,
 breaklinks=false,pdfborder={0 0 0},backref=false,colorlinks=false]
 {hyperref}

\makeatletter
\@ifundefined{textcolor}{}
{%
 \definecolor{BLACK}{gray}{0}
 \definecolor{WHITE}{gray}{1}
 \definecolor{RED}{rgb}{1,0,0}
 \definecolor{GREEN}{rgb}{0,1,0}
 \definecolor{BLUE}{rgb}{0,0,1}
 \definecolor{CYAN}{cmyk}{1,0,0,0}
 \definecolor{MAGENTA}{cmyk}{0,1,0,0}
 \definecolor{YELLOW}{cmyk}{0,0,1,0}
}

\usepackage{booktabs,array,sidecap,siunitx,nopageno}
\makeatother

\begin{document}

\title{Robust classification of salient links in complex networks}

\author{Daniel Grady}

\affiliation{Department of Engineering Sciences and Applied Mathematics, Northwestern
University, Evanston, Illinois, USA}

\author{Christian Thiemann}

\affiliation{Department of Engineering Sciences and Applied Mathematics, Northwestern
University, Evanston, Illinois, USA \& Max-Planck-Institut für Dynamik
und Selbstorganisation, Göttingen, Germany}

\author{Dirk Brockmann}

\email{brockmann@northwestern.edu}

\affiliation{Department of Engineering Sciences and Applied Mathematics \& Northwestern
Institute on Complex Systems, Northwestern University, Evanston, Illinois,
USA }
\begin{abstract}
Complex networks in natural, social, and technological systems generically
exhibit an abundance of rich information. Extracting meaningful structural
features from data is one of the most challenging tasks in network
theory. Many methods and concepts have been proposed to address this
problem such as centrality statistics, motifs, community clusters,
and backbones, but such schemes typically rely on external and arbitrary
parameters. It is unknown whether generic networks permit the classification
of elements without external intervention. Here we show that link
salience is a robust approach to classifying network elements based
on a consensus estimate of all nodes. A wide range of empirical networks
exhibit a natural, network-implicit classification of links into qualitatively
distinct groups, and the salient skeletons have generic statistical
properties. Salience also predicts essential features of contagion
phenomena on networks, and points towards a better understanding of
universal features in empirical networks that are masked by their
complexity.
\end{abstract}
\maketitle

\section{Introduction}

\global\long\def\m{\ensuremath{\#1}}
\global\long\def\p#1{\ensuremath{\mathcal{#1}}}
\global\long\def\tr#1{T^{(#1)}}
\global\long\def\spedge#1#2#3#4{\sigma(#1,#2|#3,#4)}
\global\long\def\spnode#1#2#3{\sigma_{#2#3}(#1)}
\global\long\def\ind#1{\left[#1\right]}

Many systems in physics, biology, social science, economics, and technology
are best modeled as a collection of discrete elements that interact
through an intricate, complex set of connections. Complex network
theory, a marriage of ideas and methods from statistical physics and
graph theory, has become one of the most successful frameworks for
studying these systems~\cite{Newman2003,Strogatz2001,Albert2002,Boccaletti2006,Vespignani2009,Thiemann2010,Brockmann:2010vd}
and has led to major advances in our understanding of transportation~\cite{Barrat2004a,Guimera2005a,Brockmann2006,woolley2011},
ecological systems~\cite{Allesina2008,Camacho2002}, social and communication
networks~\cite{Lazer2009}, and metabolic and gene regulatory pathways
in living cells~\cite{Jeong2000,Almaas2004,Barabasi:2004cy}.

\begin{figure*}
\centering{}\includegraphics[width=1\textwidth]{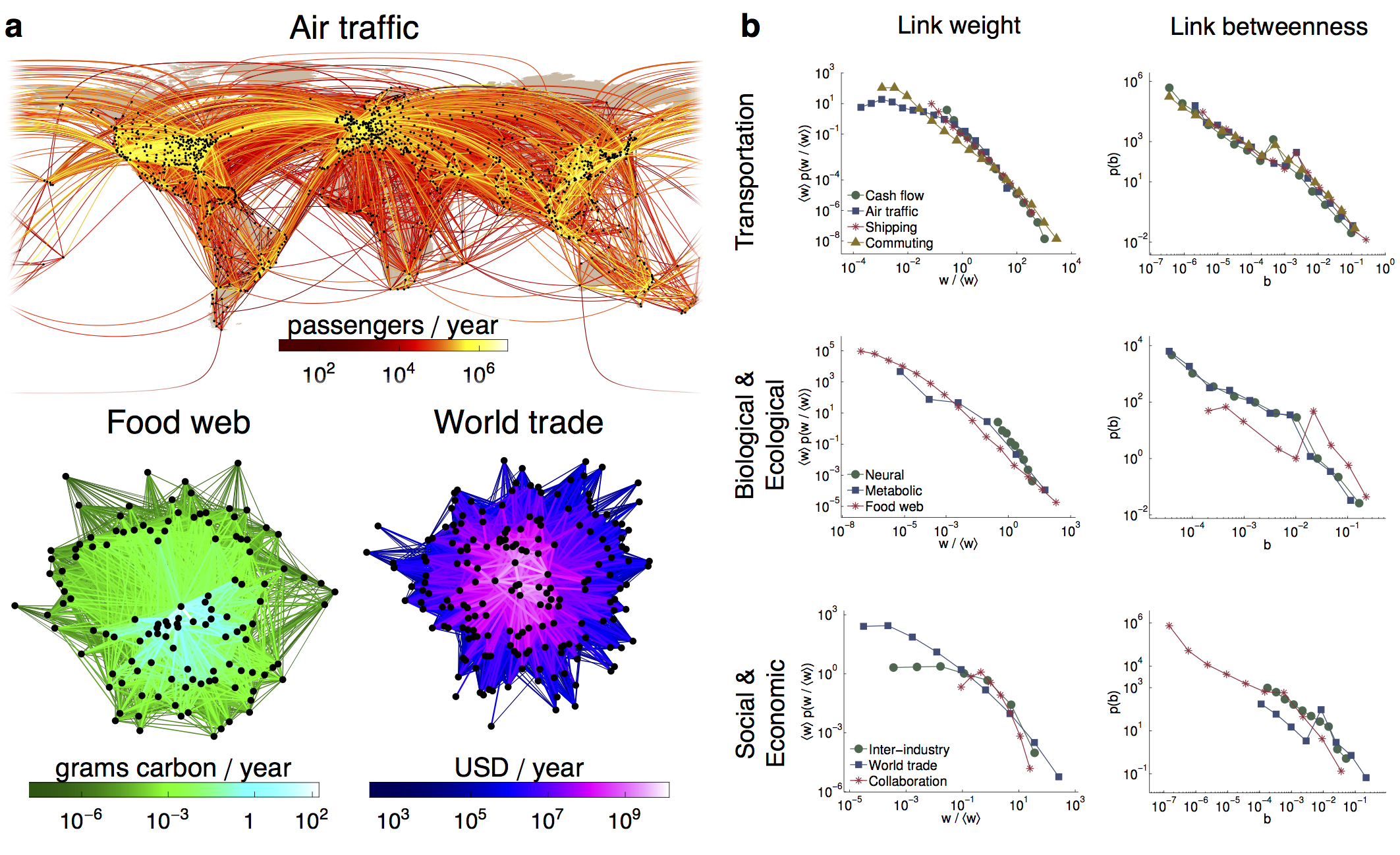}\caption{\label{Fig:empirical_networks}\textbf{Generic statistical properties
of heterogeneous complex networks.}~(\textbf{a}) Geographic representation
of the worldwide air traffic network (top), black dots represent airports,
links represent passenger flux between them, link weights $w_{ij}$
are color encoded from dark (weak) to white (strong). Networks on
the lower left and right represent the Florida bay food web and the
world trade network, respectively. Nodes in the food web are species
and links represent the exchange of biomass; in the trade network
nodes are countries and links quantify exchange in assets measured
in United States dollars (USD).~(\textbf{b}) Relative frequencies
$f(w)=\left\langle w\right\rangle p(w/\left\langle w\right\rangle )$
and $p(b)$ of link weights $w$ and link betweenness $b$ of representative
transportation, biological, ecological, social, and economic networks.
Link weights are normalized by the mean weight $\left\langle w\right\rangle $.
Details on each network are provided in Methods. In all networks link
weights and betweenness are distributed across many orders of magnitude,
and both statistics exhibit heavy tails. The substantial variability
in these quantities is also reflected in their coefficient of variation
(see Table~\ref{Tab:summary}).}
\end{figure*}

One of the challenges in complex network research is the identification
of essential structural features that are typically masked by the
network's topological complexity~\cite{Newman2003,Ravasz2003,Thiemann2010,Alon2007,Newman2004a}.
Reducing a large-scale network to its core components, filtering redundant
information, and extracting essential components are not only critical
for efficient network data management. More importantly, these methods
are often required to better understand evolutionary and dynamical
processes on networks and to identify universal principles of network
design or growth. In this context, the notion of centrality measures
according to which nodes or links can be ranked is fundamental and
epitomized by the node degree $k$, the number of directly connected
neighbors of a node. Many systems, ranging from human sexual contacts~\cite{Liljeros2001}
to computer networks~\cite{Kleinberg2001}, are characterized by
a power-law degree distribution $p(k)\sim k^{-(1+\beta)}$ with an
exponent $0<\beta\leq2$. These networks are scale-free~\cite{Barabasi1999},
meaning the majority of nodes are weakly connected and dominated by
a few strongly connected nodes, known as\emph{ hubs}. Although a variety
of networks can be understood in terms of their topological connectivity
(the set of nodes and links), a number of systems are better captured
by weighted networks in which links carry weights $w$ that quantify
their strengths~\cite{Barrat2004a,Newman2004}. An important class
of networks exhibit both a scale-free degree distribution and broadly
distributed weights which in some cases follow a power-law $p(w)\sim w^{-(1+\alpha)}$,
with $1<\alpha\leq3$~\cite{Colizza2006,Brockmann:2008p1135,Hufnagel:2004kt}.
In addition to hubs, these networks thus possess \emph{highways}.
Several representative networks of this class are depicted in Figure~\ref{Fig:empirical_networks}.
Understanding the essential underlying structures in these networks
is particularly challenging because of the mix of link and node heterogeneity.
\begin{table*}
\begin{centering}
{\scriptsize \renewcommand{\tabcolsep}{.75em} \begin{tabular}{l l r r@{.}l @{\hspace{.5em}} r@{.}l r@{.}l @{\hspace{.6em}} r@{.}l r@{.}l c r@{.}l r@{.}l @{\hspace{.5em}} r@{.}l}  & & \multicolumn{11}{c}{Full network} &  & \multicolumn{6}{c}{Salient skeleton} \\ \cmidrule{3-13} \cmidrule{15-20} Network & &   \multicolumn{1}{c}{$N$} & \multicolumn{2}{c}{$\rho$} & \multicolumn{2}{c}{$\left<k\right>$} &   \multicolumn{2}{c}{$\text{CV}(k)$} & \multicolumn{2}{c}{$\text{CV}(w)$} & \multicolumn{2}{c}{$r$} & &   \multicolumn{2}{c}{\% links} & \multicolumn{2}{c}{$\beta_\text{HSS}$} & \multicolumn{2}{c}{$r_\text{HSS}$} \\ \toprule
Cash flow       && 3,106   & 0&076               & 237&0  &  1&08  &   7&72  &  -0&137  &&   0&84  &  1&10  &  -0&255  \\ Air traffic     && 1,227   & 0&024               &  29&4  &  1&30  &   2&25  &  -0&063  &&   6&76  &  1&60  &  -0&302  \\ Shipping        && 951     & 0&057               &  54&3  &  1&22  &   7&27  &  -0&143  &&   3&66  &  1&37  &  -0&169  \\ Commuting       && 3,141   & 0&027               &  82&3  &  1&04  &  20&80  &   0&017  &&   2&44  &  2&50  &  -0&0813 \\ \midrule Neural          && 297     & 0&049               &  14&5  &  0&87  &   1&42  &  -0&163  &&  13&5   &  1&61  &  -0&308  \\ Metabolic       && 311     & 0&027               &   8&4  &  1&80  &   8&56  &  -0&253  &&  23&1   &  1&90  &  -0&381  \\ Food web        && 125     & 0&246               &  30&5  &  0&47  &  11&80  &  -0&117  &&   6&5   &  1&71  &  -0&437  \\ \midrule Inter-industry  && 128     & 1&000               & 127&0  &  0&00  &   1&70  &  -0&022  &&   1&08  &  1&58  &  -0&283  \\ World trade     && 188     & 0&446               &  83&4  &  0&65  &   8&85  &  -0&602  &&   2&39  &  1&71  &  -0&355  \\ Collaboration   && 5,835   & 8&12$\times10^{-4}$ &   4&7  &  0&96  &   1&21  &   0&185  &&  41&9   &  1&22  &  -0&242  \\
\bottomrule \end{tabular} }
\par\end{centering}{\scriptsize \par}

\caption{\label{Tab:summary}\textbf{Statistical features of the full empirical
networks and their high-salience skeletons.} Statistics for the full
networks include number of nodes $N$, link density $\rho=2L/(N^{2}-N)$
(where $L$ is the number of links), mean node degree $\left\langle k\right\rangle $,
coefficients of variation of node degree $\text{CV}(k)$ and link
weight $\text{CV}(w)$, and the assortativity coefficient $r$~\cite{Newman2002}.
For the high-salience skeletons, the first column lists the percentage
of links from the full network that are also in the HSS, an estimate
of the scaling exponent~\cite{Clauset2009} $\beta_{\text{HSS}}$
and the assortativity coefficient $r_{\text{{HSS}}}$. Further information
on network statistics are provided in Supplementary Table S1.}
\end{table*}

Although classifications of network elements according to degree,
weight, or other centrality measures have been employed in many contexts~\cite{Borgatti2006,Guimera2005a,Wu2006,Wang2008},
this approach comes with several drawbacks. The qualitative concepts
of \emph{hubs} and \emph{highways} suggest a clear-cut, network-intrinsic
categorization of elements. However, these centrality measures are
typically distributed continuously and generally do not provide a
straightforward separation of elements into qualitatively distinct
groups. At what precise degree does a node become a hub? At what strength
does a link become a highway? Despite significant advances, current
state-of-the-art methods rely on system-specific thresholds, comparisons
to null models, or imposed topological constraints~\cite{Tumminello2005,Serrano2009a,Thiemann2010,woolley2011,Radicchi2011}.
Whether generic heterogeneous networks provide a way to intrinsically
segregate elements into qualitatively distinct groups remains an open
question. In addition to this fundamental question, centrality thresholding
is particularly problematic in heterogeneous networks since key properties
of reduced networks can sensitively depend on the chosen threshold. 

Here we address these problems by introducing the concept of link
\emph{salience}. The approach is based on an ensemble of node-specific
\emph{perspectives} of the network, and quantifies the extent to which
a \emph{consensus} among nodes exists regarding the importance of
a link. We show that salience is fundamentally different from link
betweenness centrality and that it successfully classifies links into
distinct groups without external parameters or thresholds. Based on
this classification we introduce the high-salience skeleton (HSS)
of a network and compute this structure for a variety of networks
from transportation, biology, sociology, and economics. We show that
despite major differences between representative networks, the skeletons
of all networks exhibit similar statistical and topological properties
and significantly differ from alternative backbone structures such
as minimal spanning trees. Analyzing traditional random network models
we demonstrate that neither broad weight nor degree distributions
alone are sufficient to produce the patterns observed in real networks.
Furthermore, we provide evidence that the emergence of distinct link
classes is the result of the interplay of broadly-distributed node
degrees and link weights. We demonstrate how a static and deterministic
analysis of a network based on link salience can successfully predict
the behavior of dynamical processes. We conclude that the large class
of networks that exhibit broad weight and degree distributions may
evolve according to fundamentally similar rules that give rise to
similar core structures.

\section{Results}

\subsection{Link salience}

Weighted networks like those depicted in Figure~\ref{Fig:empirical_networks}
can be represented by a symmetric, weighted $N\times N$ matrix $W$
where $N$ is the number of nodes. Elements $w_{ij}\geq0$ quantify
the coupling strength between nodes $i$ and $j$. Depending on the
context, $w_{ij}$ might reflect the passenger flux between locations
in transportation networks, the synaptic strength between neurons
in a neural network, the value of assets exchanged between firms in
a trade network, or the contact rate between individuals in a social
network.

Our analysis is based on the concept of \emph{effective proximity}
$d_{ij}$ defined by the reciprocal coupling strength $d_{ij}=1/w_{ij}$.
Effective proximity captures the intuitive notion that strongly (weakly)
coupled nodes are close to (distant from) each other~\cite{Caldarelli2007}.
It also provides one way to define the length of a path $\p P$ that
connects two terminal nodes ($n_{1},n_{K}$) and consists of $K-1$
legs via a sequence of intermediate nodes $n_{i}$, and connections
$w_{n_{i}n_{i+1}}>0$. The \emph{shortest path} minimizes the total
\emph{effective distance} $l=\sum_{i=1}^{K-1}d_{n_{i}n_{i+1}}$ and
can be interpreted as the most efficient route between its terminal
nodes~\cite{Dijkstra1959a,Newman2010}; this definition of shortest
path is used throughout this paper. In networks with homogeneous weights,
shortest paths are typically degenerate, and many different shortest
paths coexist for a given pair of terminal nodes. In heterogeneous
networks with real-valued weights shortest paths are typically unique.
For a fixed reference node $r$, the collection of shortest paths
to all other nodes defines the shortest-path tree (SPT) $T(r)$ which
summarizes the most effective routes from the reference node $r$
to the rest of the network. $T(r)$ is conveniently represented by
a symmetric $N\times N$ matrix with elements $t_{ij}(r)=1$ if the
link $(i,j)$ is part of at least one of the shortest paths and $t_{ij}(r)=0$
if it is not.

The central idea of our approach is based on the notion of the \emph{average}
\emph{shortest-path tree} as illustrated in Figure~\ref{Fig:salience}a.
We define the salience $S$ of a network as 
\begin{equation}
S=\left\langle T\right\rangle =\frac{1}{N}\sum_{k}T(k)\label{eqn:salience}
\end{equation}
so that $S$ is a linear superposition of all SPTs. $S$ can be calculated
efficiently using a variant of a standard algorithm (see Supplementary
Methods). According to this definition the element $0\leq s_{ij}\leq1$
of the matrix $S$ quantifies the fraction of SPTs the link $(i,j)$
participates in. Since $T(r)$ reflects the set of most efficient
paths to the rest of the network from the perspective of the reference
node, $s_{ij}$ is a \emph{consensus variable} defined by the ensemble
of root nodes. If $s_{ij}=1$ then link $(i,j)$ is essential for
all reference nodes, if $s_{ij}=0$ the link plays no role and if,
say, $s_{ij}=1/2$ then link $(i,j)$ is important for only half the
root nodes. Note that although $S$ is defined as an average across
the set of shortest-path trees, it is itself not necessarily a tree
and is typically different from known structures such as minimal spanning
trees (see Supplementary Figure S1, Supplementary Table S3 and Supplementary
Methods).

\subsection{Robust classification of links}

\begin{figure*}
\includegraphics[width=1\textwidth]{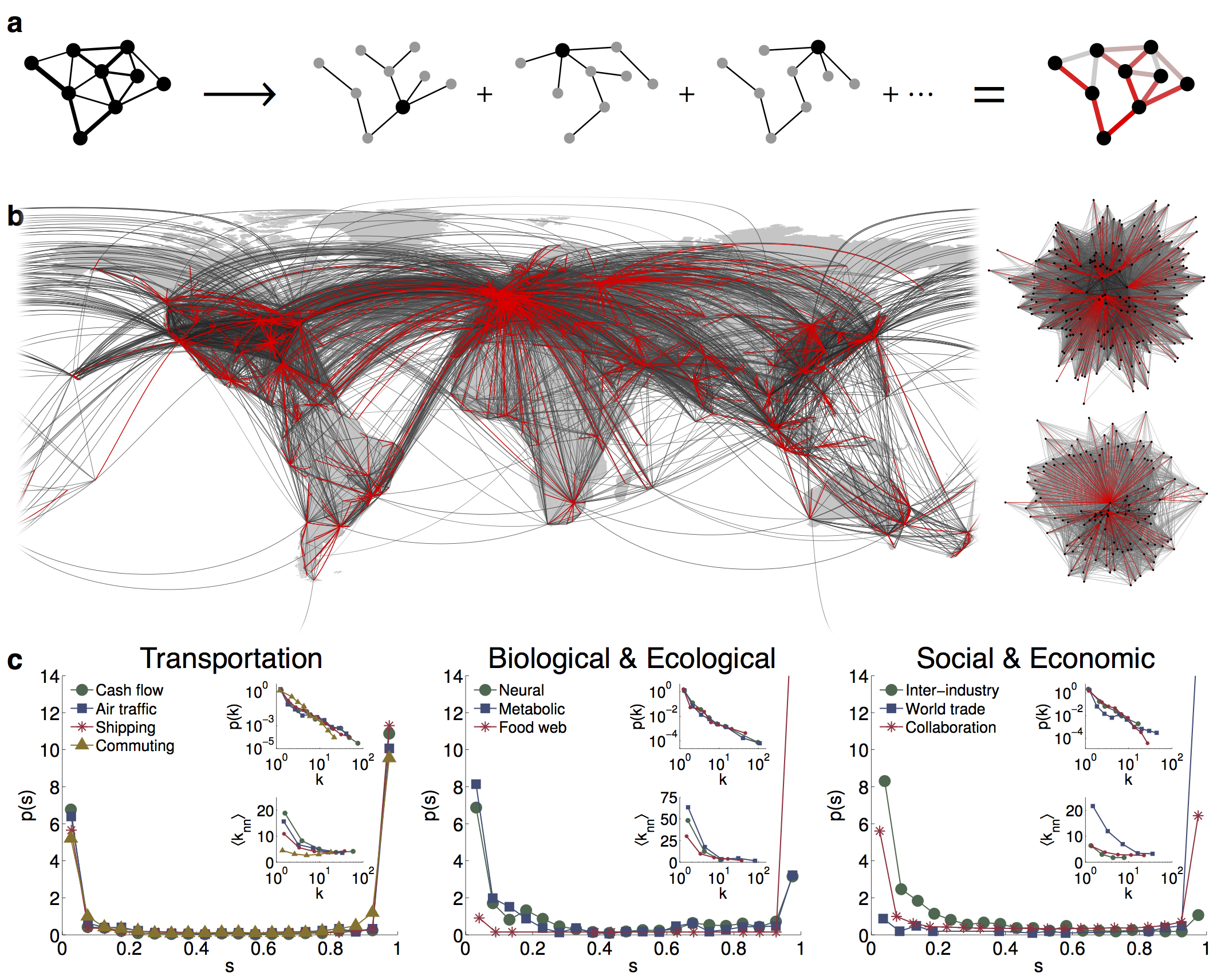}\caption{\label{Fig:salience}\textbf{Computation of link salience and properties
of the high-salience skeleton.}~(\textbf{a}) For each reference node
$r$ in the weighted network on the left the shortest-path tree $T(r)$
is computed. The superposition of all trees according to Equation~(\ref{eqn:salience})
assigns a value $s_{ij}$ to each link in the original network. Salience
values are shown on the right with link color: red is high salience
and grey is low.~(\textbf{b}) The collection of high salience links
(red) for the networks shown in Figure~\ref{Fig:empirical_networks}.
The full networks are shown in grey.~(\textbf{c}) The relative frequency
$p(s)$ of non-zero salience values $s$. The distribution $p(s)$
is bimodal in all networks under consideration. This key feature of
bimodality of $p(s)$ provides a plausible, parameter-insensitive
classification of links, salient ($s\approx1$) vs. non-salient ($s\approx0$),
and implies that nodes in these networks typically \emph{agree} whether
a link is essential or not. The high-salience skeleton (HSS) is defined
as the collection of links that accumulate near $s\approx1$. Upper
and lower insets depict, respectively, the degree distribution $p(k)$
of the HSSs and mean next-neighbor degree $\left\langle k_{nn}|k\right\rangle $
as a function of node degree $k$. The HSS degree distribution is
typically scale-free (see Supplementary Figure S2) and the skeletons
are typically strongly disassortative. Note that although they may
be, and often are, divided into multiple components, the largest connected
component of the skeleton typically dominates. This connectedness
is not imposed, but is an emergent property of salience. (See Supplementary
Table S2).}
\end{figure*}
The most important and surprising feature of link salience is depicted
in Figure~\ref{Fig:salience}c. For the representative set of networks,
we find that the distribution $p(s)$ of link salience exhibits a
characteristic bimodal shape on the unit interval. The networks' links
naturally accumulate at the range boundaries with a vanishing fraction
at intermediate values. Salience thus successfully classifies network
links into two groups: salient ($s\approx1$) or non-salient ($s\approx0$),
and the large majority of nodes \emph{agree} on the importance of
a given link. Since essentially no links fall into the intermediate
regime, the resulting classification is insensitive to an imposed
threshold, and is an intrinsic and emergent network property characteristic
of a variety of strongly heterogeneous networks. This is fundamentally
different from common link centrality measures such as weight or betweenness
that possess broad distributions (see Fig.~\ref{Fig:empirical_networks}b),
and which require external and often arbitrary threshold parameters
for meaningful classifications~\cite{Serrano2009a,Radicchi2011}.

The salience as defined by Eq.~\ref{eqn:salience} permits an intuitive
definition of a network's skeleton as a structure which incorporates
the collection of links that accumulate at $s\approx1$. Figure~\ref{Fig:salience}b
depicts the skeleton for the networks of Figure~\ref{Fig:empirical_networks}a.
For all networks considered, only a small fraction of links are part
of the high-salience skeleton (6.76\% for the air traffic network,
6.5\% for the food web, and 2.39\% for the world trade network), and
the topological properties of these skeletons are remarkably generic.
Note that technically a separation of links into groups according
to salience requires the definition of a threshold (e.g. we chose
the center of the salience range for convenience). The important feature
is that the resulting groups are robust against changes in the value,
since almost no links fall into intermediate ranges. Consequently
the point of separation is almost arbitrary, yield almost identical
skeletons for threshold ranges of $80\%$ of the entire range. One
of the common features of these skeletons is their strong disassortativity,
irrespective of the assortativity properties of the corresponding
original network~(see Table~\ref{Tab:summary}). Furthermore, all
skeletons exhibit a scale-free degree distribution 
\begin{equation}
p_{\text{HSS}}(k)\sim k^{-(1+\beta_{\text{HSS}})}\label{eq:schnickschnack}
\end{equation}
with exponents $1.1\leq\beta_{\text{HSS}}\leq2.5$ (see Table~\ref{Tab:summary}
and Supplementary Figure S2). Since only links with $s\approx1$ are
present in the HSS, the degree of a node in the skeleton can be interpreted
as the total salience of the node. The collapse onto a common scale-free
topology is particularly striking since the original networks range
from quasi-planar topologies with small local connectivity (the commuter
network) to completely connected networks (worldwide trade). Note
that the lowest exponent (weakest tail) is observed for the commuter
network, since in a quasi-planar network the maximum number of salient
connections is limited by the comparatively small degree of the original
network. The scale-free structure of the HSS consequently suggests
that networks that possess very different statistical and topological
properties and that have evolved in a variety of contexts seem to
self-organize into structures that possess a robust, disassortative
backbone, despite their typical link redundancy. 

Although these properties of link salience are encouraging and suggest
novel opportunities for filtering links in complex weighted networks,
for understanding hidden core sub-structures, and suggest a new mechanism
for defining a network's skeleton, a number of questions need to be
addressed and clarified in order for the approach to be viable. First,
a possible criticism concerns the definition of salience from shortest-path
trees which suggests that $s_{ij}$ can be trivially obtained from
link betweenness $b_{ij}$, for example by means of a non-linear transform.
Secondly, a bimodal $p(s)$ may be a trivial consequence of broad
weight distributions, if for instance large weights are typically
those with $s\approx1$. Finally, the observed bimodal shape of $p(s)$
could be a property of any non-trivial network topology such as simple
random weighted networks. In the following we will address each of
these concerns.

\subsection{Salience and betweenness\label{sub:Relationship-to-betweenness}}

\begin{SCfigure*}
\includegraphics[width=\columnwidth]{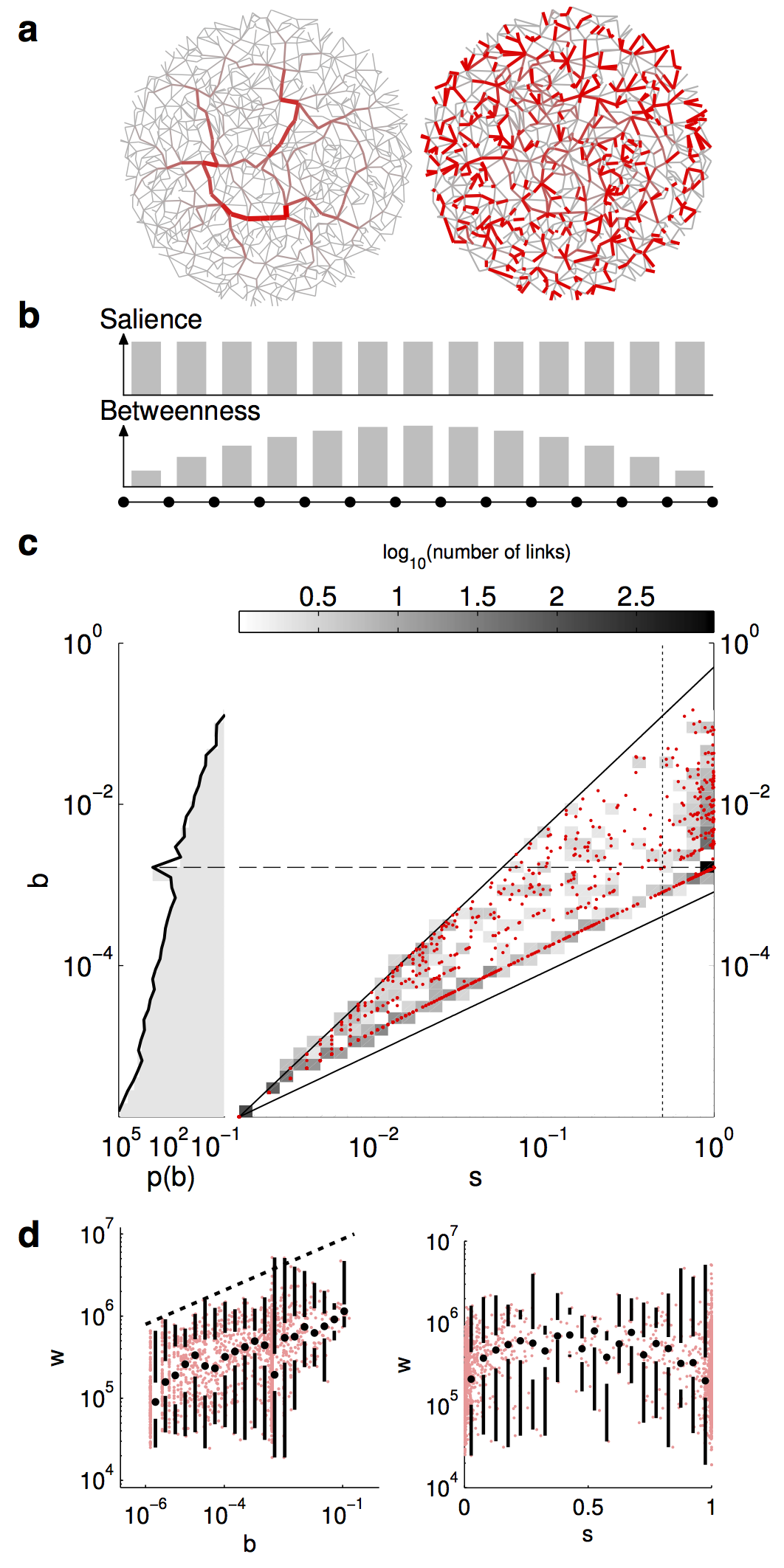}
\caption{\label{Fig:bvs}\textbf{Salience and betweenness capture different aspects of centrality.}~ (\textbf{a}) A schematic planar network in which the color of links quantifies betweenness $b$ (left) and salience $s$ (right). High-betweenness links tend to be located near the barycenter of the network~\cite{Barthelemy2011}, whereas high-salience links are distributed evenly throughout the network.~ (\textbf{b}) A simple linear chain shows the reason for this effect. A link in the center serves as a shortest-path bridge between all pairs of nodes, and so has the highest betweenness. But since all shortest-path trees are identical, all links have identical salience.~ (\textbf{c}) A scatter plot (red dots) of link salience $s$ versus link betweenness $b$ for the air traffic network (point density is quantified in grey). The vertical dotted line marks $s=1/2$ and the solid curves represent the theoretical bounds of Equation~(\ref{eq:wurst}). The projected density $p(b)$ is shown on the left. The lack of any clear correlation in the scatter plot is typical of all networks in Figure~\ref{Fig:empirical_networks}. (See Supplementary Figure S3 for additional correlograms.)~ (\textbf{d}) Scatter plots (in light red) of betweenness $b$ (left) and salience $s$ (right) versus link weight $w$ in the air traffic network. The bottom and top of the lower whiskers, the dot, and the bottom and top of the upper whiskers correspond to the 0, 25, 50, 75, and 100th percentiles, respectively. The dashed line indicates a scaling relationship $w\sim b^{\gamma}$ with $\gamma\approx0.2$. Although the network exhibits a positive correlation between link weight and link betweenness, the high-salience skeleton incorporates links with weights spanning the entire range of observed values; no clear correlation of weight with salience exists. These properties are observed in the other networks as well.}
\end{SCfigure*} The betweenness $b_{ij}$ of a link $(i,j)$ is the fraction of all~$\sim N^{2}$
shortest paths that pass though the link, whereas the salience $s_{ij}$
is the fraction of $N$ shortest-path trees $T(r)$ the link is part
of. Despite the apparent similarity between these two definitions,
both quantities capture very different qualities of links, as illustrated
in Figure~\ref{Fig:bvs}. Betweenness is a centrality measure in
the traditional sense~\cite{Freeman1977}, and is affected by the
topological position of a link. Networks often exhibit a core-periphery
structure~\cite{Holme2005} and the betweenness measure assigns greater
weight to links that are closer to the barycenter of the network~\cite{Barthelemy2011}.
Salience, on the other hand, is insensitive to a link's position,
acting as a uniform filter. This is illustrated schematically in the
random planar network of Figure~\ref{Fig:bvs}a. High betweenness
links tend to be located in the center of the planar disk, whereas
high salience links are distributed uniformly. A given shortest path
is more likely to cross the center of the disk, whereas the links
of a shortest-path tree are uniformly distributed, as they have to
span the full network by definition. A detailed mathematical comparison
of betweenness and salience is provided in the Methods. Figure~\ref{Fig:bvs}c
depicts the typical relation of betweenness and salience in a correlogram
for the worldwide air traffic network. The data cloud is broadly distributed
within the range of possible values given by the inequalities (see
Supplementary Methods)
\begin{equation}
s/N\leq b\leq s^{2}/2.\label{eq:wurst}
\end{equation}
Within these bounds no functional relationship between $b$ and $s$
exists. Given a link's betweenness $b$ one generally cannot predict
its salience and vice versa. In particular, high-salience links ($s\approx1$)
possess betweenness values ranging over many scales. The spread of
data points within the theoretical bounds is typical for all the networks
considered (see Supplementary Figure S3). Links tend to collect at
the right-hand edge, corresponding to the upper peak in salience,
and in particular at the lower right corner of the wedge-shaped region,
corresponding to the heretofore-unexplained peak in betweenness exhibited
by several of the networks (cf.~Figure~\ref{Fig:empirical_networks}
and the dashed line in Figure~\ref{Fig:bvs}b). These edges have
maximal salience (all nodes agree on their importance) but the smallest
betweenness possible given this restriction (they are not well-represented
in the set of shortest paths). Such edges are the spokes in the hub-and-spoke
structure: they connect a single node to the rest of the network,
but are used by no others, and they are an essential piece of the
high-salience skeleton, since severing them removes some node's best
link to the main body of the network. The presence of such links in
the high-salience skeleton explains why the weight values of $s\approx1$
edges span such a wide range, since a link may have relatively low
weight and yet be some node's most important connection.

Figure~\ref{Fig:bvs}d tests the hypothesis that strong link weights
may yield strong values for salience. We observe that link betweenness
is positively correlated with link weight and roughly follows a scaling
relation $w\sim b^{\gamma}$ with $\gamma\approx0.2$, in agreement
with previous work on node centrality~\cite{DallAsta2006}. This
is not surprising since high-weight links are by definition shorter
and tend to attract shortest paths. In contrast, link weights exhibit
no systematic dependence on salience, and in particular large weights
do not generally imply large salience. In fact, for fixed link salience
the distribution of weights is broad with approximately the same median.
Consequently, salience can be considered an independent centrality
dimension that measures different features than correlated centrality
measures such as weight and betweenness.

\subsection{Origin of bimodal salience}

\begin{figure}
\includegraphics[width=1\columnwidth]{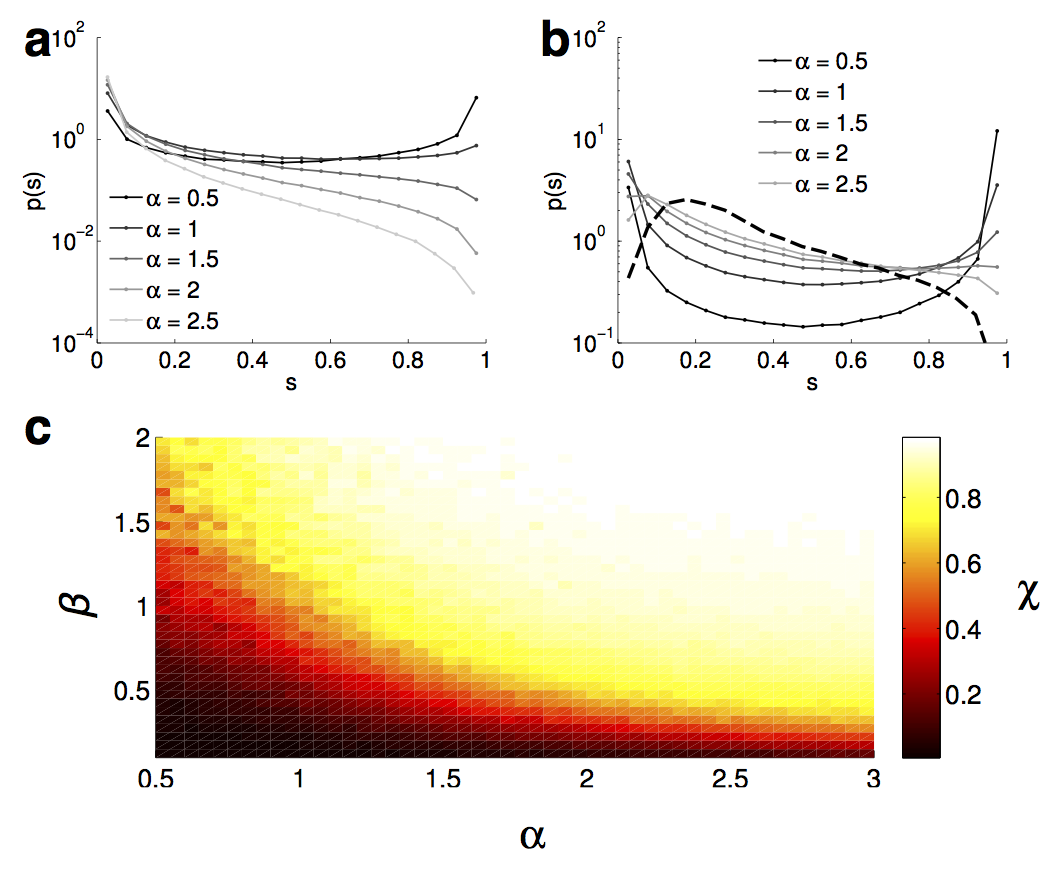}\caption{\label{fig4}\textbf{Salience in random networks.}~ (\textbf{a})
Salience distributions $p(s)$ in fully connected networks with 1,000
nodes and weights assigned using a power law $p(w)\sim w^{-(1+\alpha)}$
for various tail exponents $\alpha$. Complete networks serve as models
of systems with all-to-all interactions, such as the inter-industry
trade network. Only for unrealistically broad weight distributions
($\alpha\lesssim1$) does $p(s)$ exhibit a bimodal character. If
$\alpha>1$ bimodality is absent.~ (\textbf{b}) Salience distributions
in preferential attachment networks (1,000 nodes)~\cite{Barabasi1999}
with degree distribution $p(k)\sim k^{-3}$ and uniform weights do
not exhibit bimodal salience (heavy dashed line). If however the power-law
weight distribution is superimposed on the preferential attachment
topology, bimodal salience emerges for realistic values of $\alpha$.~
(\textbf{c}) For the range of tail exponents $\alpha$ and $\beta$
the color code quantifies the magnitude $\chi$ of bimodality in the
salience distribution $p_{\alpha.\beta}(s)$ of a network with a scale-free
degree distribution with exponent $\beta$ (constructed using the
configuration model~\cite{Boccaletti2006}) and a scale-free weight
distribution with exponent $\alpha$. Small values of $\chi$ correspond
to a bimodal $p_{\alpha,\beta}(s)$. The bimodality measure $\chi$
was computed using Kolmogorov-Smirnov distance between $p_{\alpha,\beta}(s)$
for $s>0$ and the idealized reference distribution $q(s)=\delta(s-1)$.}
\end{figure}
All the networks we consider feature broad link weight distributions
$p(w)$ (see Figure~\ref{Fig:empirical_networks}b), some of which
can be reasonably modeled by power laws $p(w)\sim w^{-(\alpha+1)}$
with exponents for many empirical data sets typically in the range
$1<\alpha<3$~\cite{Clauset2009} (smaller $\alpha$ corresponds
to broader $p(w)$). Although it may seem plausible that strong links
in the tail of these distributions dominate the structure of shortest-path
trees and thus cause the characteristic bimodal distribution of link
salience, evidence against this hypothesis is already apparent in
Fig~\ref{Fig:bvs}d: links with high salience exhibit weights across
many scales, and in particular low-weight links may possess high salience.
Further evidence is provided in Figure~\ref{fig4}a, which depicts
the salience distribution for fully connected networks for a sequence
of tail parameters $\alpha$. For values of $\alpha$ in the range
observed in real networks, $p(s)$ is peaked near $s=0$ and decreases
with increasing $s$. A bimodal distribution of $s$ only emerges
when $\alpha$ is unrealistically small ($\alpha<1$), and is much
less pronounced than in real networks (cf.~Figure~\ref{Fig:salience}).
We conclude that broad, scale-free weight distributions $p(w)$ alone
are insufficient to cause the natural, bimodal distribution $p(s)$
observed in real networks.

Another potential source of the observed bimodality in $p(s)$ is
the \emph{topological heterogeneity} of a scale-free degree distribution
$p(k)\sim k^{-(1+\beta)}$ with $0<\beta<2$~\cite{Barabasi1999,Kleinberg2001,Guimera2007}.
Figure~\ref{fig4}b provides evidence that also a scale-free topology
alone does not yield the characteristic bimodal salience distribution.
In fact, the generic preferential attachment network~\cite{Barabasi1999}
($\beta=2$) with uniform weights exhibits a distribution of salience
that is almost the complement of the observed pattern with mostly
intermediate values of link salience. The presence of hubs implies
that any shortest paths seeking out a node in a hub's region will
most likely route through that hub, and links emanating from this
hub are more likely to appear in many shortest-path trees. However,
the hub-and-spoke structure of a preferential attachment network is
only approximate; nodes that are at the end of a spoke are still likely
to have random links to other areas of the network. For this reason,
it is not typical in the uniform-weight preferential attachment network
to find links that appear in nearly all shortest-path trees.

However, the observed bimodal distribution $p(s)$ can be generated
in random networks by a combination of weight and degree variability,
a property characteristic of the class of networks discussed here.
Figure~\ref{fig4}b also depicts $p(s)$ for preferential attachment
networks that possess a scale-free distribution of both degree $k$
and weight $w$. As the weight distribution becomes broader (decreasing
$\alpha$), and even in the absence of explicit degree-weight correlations,
we see the emergence of bimodality in the salience distribution in
these networks. Topological hubs are more likely to have extremely
high-weight links simply because they have more links. Even when there
is a topologically short path terminating at a spoke node that does
not pass through the corresponding hub, it is less likely to be the
shortest weighted path. Extreme weights amplify the effects of hubs
by drawing more shortest paths through them. Moreover, Figure~\ref{fig4}c
demonstrates that the emergence of bimodal salience does depend on
the interplay between degree and weight distributions: the broader
the degree distribution, the narrower the required weight distribution.

All of these results support the conclusion that a bimodal salience
distribution is characteristic of networks with strong heterogeneity
in both topology and interaction strength, but that unweighted networks
do not exhibit this property.

\subsection{Applications to network dynamical systems\label{sec:infection}}

\begin{figure}
\includegraphics[width=1\columnwidth]{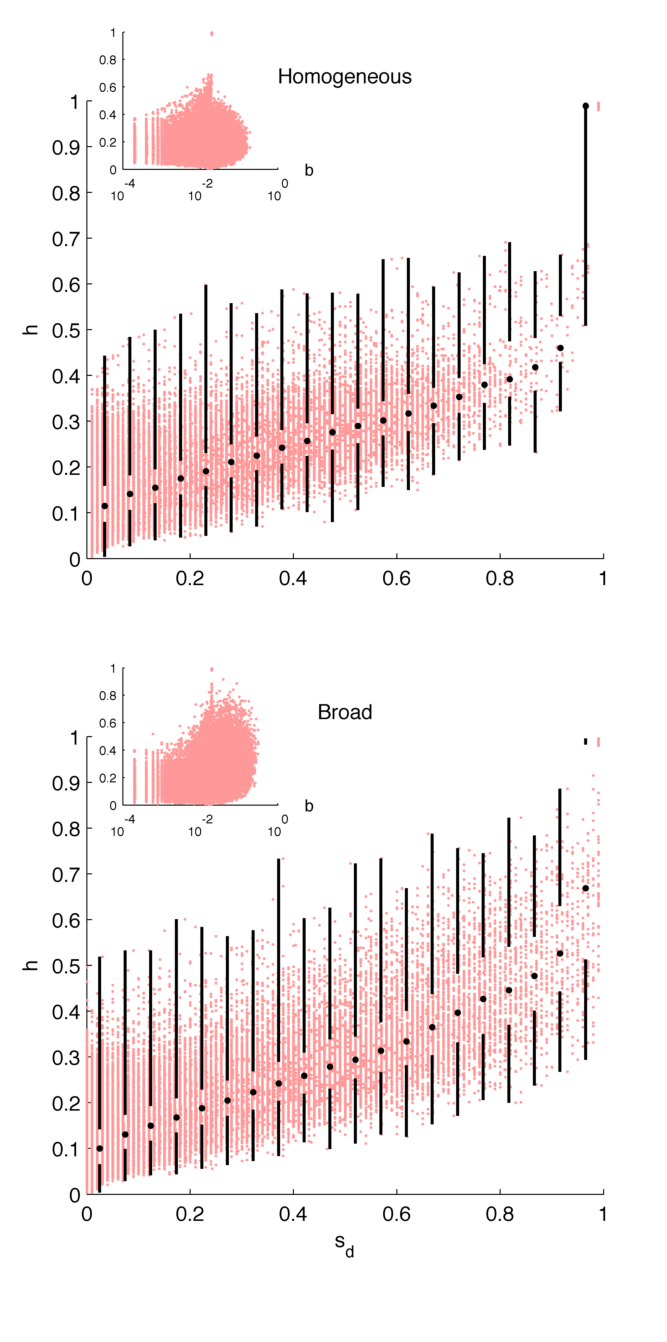}\caption{\label{fig:Salience-infection}\textbf{Salience predicts infection
pathways in stochastic epidemic models.} The scatter plots show the
directed salience $s_{d}$ against the normalized frequency of appearance
in infection pathways $h$ for each link in an ensemble of 100 networks,
averaged over 1,000 epidemic realizations for each member of the ensemble.
As in Figure~\ref{Fig:bvs}, the plots are divided horizontally into
bins, with the heavy black lines indicating quartiles within each
bin. Insets show link betweenness $b$ versus $h$, and correlation
coefficients are listed in Table~\ref{tab:salience-infection}.~
\textbf{top}, Weights distributed narrowly and uniformly around a
constant $w_{0}$.~ \textbf{bottom}, Weights distributed according
to $p(w)\sim w^{-(1+\alpha)}$ with $\alpha=2$.}
\end{figure}
The relevance of link salience to dynamical processes that evolve
on networks is an important issue, and one area of particular interest
in network research is contagion phenomena. In this context, individuals
in a population are represented by nodes, and interaction propensities
between pairs of nodes by a weighted network. Contagion phenomena
are modeled by transmissions between nodes along the links of the
network, where the likelihood of transmission is quantified by the
link weights. The central question in this class of models is how
the topological properties of the network shape the dynamics of the
process. Link salience can also provide useful information about the
behavior of such a dynamical system. To illustrate this, we consider
a simple stochastic SI epidemic model. At any given point in time,
an infected node $i$ can transmit a disease to susceptible nodes
at a rate determined by the link weight $w_{ji}$. The details of
the model are provided in the Methods. We consider an epidemic on
a planar disk network similar to that shown in Figure~\ref{Fig:bvs}a.
A single node is chosen at random for the outbreak location. At every
step of the process each infected node randomly selects a neighbor
to infect with probability proportional to the link weight; eventually
the entire network is infected. By keeping track of which links were
used in the infection process one obtains the infection hierarchy
$H$, a directed tree structure that represents the epidemic pathway
through the network. Since the process is stochastic, each realization
of the process generates a different infection hierarchy. For different
initial outbreak nodes and realizations of the process we calculate
an infection frequency $h$ for each link: The number of times that
link is used in the infection process, normalized by the number of
realizations. The question is, how successfully can link salience,
a topological quantity, predict infection frequency $h$, a dynamic
quantity. Figure~\ref{fig:Salience-infection} shows the results
for the two different link weight scenarios described in the Methods.
The top panel shows networks with link weights narrowly and uniformly
distributed around a constant value $w_{0}$; in the bottom panel
link weights are broadly distributed according to a power law. In
both cases, link salience is highly correlated with the frequency
of a link's appearance in infection hierarchies $h$, while alternative
link centrality measures such as weight and betweenness are not (see
Figure~\ref{fig:Salience-infection} insets and SI). The link salience
on average gives a much more accurate prediction of the virulence
of a link than other available measures of centrality, suggesting
that this type of completely deterministic, static analysis could
nonetheless play an important role in considering how best to slow
spreading processes in real networks.

\section{Discussion}

As much recent work in network theory has shown~\cite{Serrano2009a,Tumminello2005,Newman2004a,Alon2007},
there is tremendous potential for extracting heretofore hidden information
from the complex interactions between the elements of a system. However,
until now these methods have relied on externally imposed parameters
or null models. Here we have shown that typical empirical networks
taken from a variety of fields do in fact permit the robust classification
of links according to the node-consensus procedure we introduce, and
that this leads naturally to the definition of a high-salience skeleton
in these networks. Because vanishingly few links in empirical networks
have intermediate values of salience, the identification of the skeleton
is insensitive to a salience threshold; indeed, if a tunable filtering
procedure is desired other methods may be more appropriate. Not all
networks possess a skeleton; simple unweighted models have a shortest-path
structure spread throughout the links. However, the presence of a
skeleton is a generic feature of many heterogeneously weighted, empirical
networks. We suggest that the likely cause in real networks is a hub-and-spoke
topological structure along with a broad weight distribution, which
amplifies the tendency of hubs to capture shortest paths.

We believe that the concept of salience and the high-salience skeleton
will become a vital component in understanding networks of the type
discussed here and the development of network-based dynamical models.
The simple SI model we investigate here is only a starting point;
it may be possible to leverage knowledge of a network's high-salience
skeleton to develop dynamical models that do not require simulation
on (or even knowledge of) the full network. The generic bimodal salience
distribution in this context also implies that in contagion phenomena
only a small subset of links might typically be active even if the
process is stochastic. Those links, however, are almost certainly
active irrespective of the outbreak location and the stochasticity
of the process, which implies that in this regime the process becomes
more predictable and the impact of stochasticity is decreased. This
effect may shed a new light on the impact of stochastic factors in
disease dynamical processes that evolve in strongly heterogeneous
networks.

Many of the networks we considered evolved over long periods of time
subject to external constraints and unknown optimization principles.
The discovery that pronounced weight and degree heterogeneity, which
are defining properties of the investigated networks, go hand in hand
with generic properties in their underlying skeleton indicate that
looking for common evolution principles could be another promising
direction of further research.

\section{Methods}

\subsection{Network data sources}

\begin{table*}
\begin{centering}
\begin{tabular}{lll} Network        & Nodes                               & Link units                                    \\ \toprule Cash flow      & Counties, continental United States & Number of bills/time                          \\ Air traffic    & Airports, worldwide                 & Number of passengers/time                     \\ Shipping       & Ports, worldwide                    & Number of cargo ships/time                    \\ Commuting      & Counties, continental United States & Number of commuters/times                     \\ \midrule Neural         & Neurons, \emph{C. elegans}          & Number of synapses and gap junctions          \\ Metabolic      & Metabolites, \emph{E. coli}         & Effective kinetic reaction rate               \\ Food web       & Species, Florida Bay food web       & Exchanged biomass/time                        \\ \midrule Inter-industry & Industrial sectors, United States   & Average input required for fixed output (USD) \\ World trade    & Countries                           & Average value of traded assets/time (USD)     \\ Collaboration  & Scientists                          & Number of co-authored papers                  \\ \bottomrule \end{tabular} 
\par\end{centering}

\caption{\label{Tab:descriptions}\textbf{Definition of nodes and links in
empirical networks. }The entities represented by nodes, as well as
the units measured by link weight, are listed for every network.}
\end{table*}
Table~\ref{Tab:descriptions} gives a brief definition of each network
we examine here, and below we provide a summary of the networks along
with data sources and references.

The Cash flow network was constructed from data collected through
the Where's George bill-tracking website (\url{http://www.wheresgeorge.com}).
The nodes are the 3,106 counties in the 48 United States excluding
Alaska and Hawaii, and the links measure the number of bills passing
between pairs of counties per time. This network has been previously
analyzed~\cite{Brockmann2006,Thiemann2010,Brockmann:2008p1135};
see in particular the supplement to~\cite{Thiemann2010} for a wealth
of detailed information regarding the construction and statistics
of this network, as well as strong evidence for interpreting it as
proxy for individual mobility. The network of cash flow is constructed
from approximately 10 million individual bank notes that circulate
in the United States.

The Air traffic network measures global air traffic based on flight
data provided by OAG Worldwide Ltd. (\url{http://www.oag.com}) and
includes all scheduled commercial flights in the world. Nodes represent
airports worldwide. Link weights measures the total number of passengers
traveling between a pair of networks by direct flights per year. This
network is well-represented in the literature~\cite{Barrat2004a,Colizza2006,Guimera2005a,Guimera2007,Sales-Pardo2007};
we reduce it to 95\% flux as described in \cite{Olivia1998}. Total
traffic in this network amounts to approximately 3 billion passengers
per year.

The Shipping network quantifies international marine freight traffic
based on data provided by IHS Fairplay (\url{http://www.ihs.com/products/maritime-information/index.aspx})
which includes itineraries for 16,363 container ships. Nodes represent
ports, and links measure the number of commercial cargo vessels traveling
between those ports during 2007. The network is available at \url{http://www.mathmod.icbm.de/45365.html}
and further discussion can be found in~\cite{Kaluza2010}.

The Commuting network is based on surveys conducted by the US Census
Bureau during the 2000 census, and reflects the daily commuter traffic
between US counties; the data is publicly available at \url{http://www.census.gov/population/www/cen2000/commuting/files/2KRESCO_US.zip}.
Nodes in this network represent the counties of the 48 states excluding
Alaska and Hawaii, and links measure the number of people commuting
between pairs of counties per day.

The Neural network is derived from the \emph{Caenorhabditis elegans}
nematode. Nodes represent neurons, and links measure the number of
synapses or gap junctions connecting a pair of neurons. Experimental
data is described in Ref.~\cite{White1986} and analyzed in Ref.~\cite{Watts1998};
the network is available at \url{http://www-personal.umich.edu/~mejn/netdata/}.

The metabolic network measures interactions in the bacterium \emph{Escherichia
coli}~\cite{Reed2003,Almaas2004}. Nodes represent metabolites and
links measure effective kinetic rates of reactions a pair of metabolites
participates in. We use only the largest connected component of this
network.

The Food web network is a representative food web from a list of publicly
available data sets of the same type (see \url{http://vlado.fmf.uni-lj.si/pub/networks/data/bio/foodweb/foodweb.htm}
for networks in Pajek format, a report~\cite{Ulanowicz1998} on trophic
analysis of the Florida Bay food web available at \url{http://www.cbl.umces.edu/~atlss/FBay701.html},
and Refs.~\cite{Serrano2009a,Radicchi2011}). Nodes represent species
in the Florida Bay ecosystem, and links measure the consumed biomass
in grams of carbon per year across a link.

In the Inter-industry network, nodes represent industrial sectors
in the United States and their connections are computed from input-output
tables prepared by the US Bureau of Economic Analysis available at
\url{http://www.bea.gov/industry/io_benchmark.htm}. We use data from
2002, the most recent year for which measurements are available. Nodes
in this network represent particular industries (for example, ``tobacco
production'' or ``cutlery and hand tool manufacturing'') and links
measure an average interaction between two industries. Given two industries
$x$ and $y$, input-output data measures the amount (USD) of input
$x$ demands from $y$ in order to produce one dollar of output, and
we take the weight of the link connecting $x$ and $y$ to be the
geometric mean of the input-output demand of $x$ on $y$ and $y$
on $x$.

The World trade network is based on data prepared by the United States
National Bureau of Economic Research and measures the value (in nominal
thousands of USD) of goods traded between countries from 1962-2000.
Nodes represent countries and links measure the value of goods traded
between countries. The data and extensive documentation are available
at \url{http://cid.econ.ucdavis.edu/data/undata/undata.html}. A series
of papers analyzes a similar data set from a different source~\cite{Radicchi2011,Garlaschelli2004,Garlaschelli2005,Serrano2003}.

The Collaboration network is based on co-authorship of academic papers
in the high-energy physics community from 1995-1999. Nodes represent
individuals and links measure the number of papers co-authored~\cite{Newman2001a}.
The data is publicly available at \url{http://www-personal.umich.edu/~mejn/netdata/}.

\subsection{Link salience and betweenness centrality}

Link salience $s$ and betweenness centrality $b$ are based on the
notion of shortest paths in weighted networks. Given a weighted network
defined by the weight matrix $w_{ij}$ (not necessarily symmetric)
and a shortest path that originates at node $x$ and terminates at
node $y$ it is convenient to define the indicator function 
\[
\sigma_{ij}(y,x)=\begin{cases}
1 & \text{if link }i\rightarrow j\text{ is on the shortest path}\\
 & \quad\text{from }x\text{ to }y\\
0 & \text{otherwise}
\end{cases}
\]
A shortest path tree $T(x)$ rooted at node $x$ can be represented
as a matrix with elements 
\[
T_{ij}(x)=\begin{cases}
1 & \text{if }\sum_{y}\sigma_{ij}(y,x)>0\\
0 & \text{otherwise},
\end{cases}
\]
and salience $s_{ij}$ of link $i\rightarrow j$ is given by
\begin{equation}
s_{ij}=\frac{1}{N}\sum_{x}T_{ij}(x)=\left\langle T_{ij}(x)\right\rangle _{V}\label{eq:marmelade}
\end{equation}
where $\left\langle \cdot\right\rangle _{V}$ denotes the average
across the set of root nodes $x$. 

Betweenness, on the other hand, is defined according to
\[
b_{ij}=\frac{1}{N^{2}}\sum_{x,y}\sigma_{ij}(y,x)=\left\langle \sigma_{ij}(y,x)\right\rangle _{V^{2}}
\]
where $\left\langle \cdot\right\rangle _{V^{2}}$ denotes the average
over all $N^{2}$ pairs of terminal nodes. The relation of betweenness
and salience can be made more transparent by rewriting this expectation
value as a sequential average over all nodes,
\[
b_{ij}=\frac{1}{N}\sum_{x}b_{ij}(x)
\]
with
\[
b_{ij}(x)=\frac{1}{N}\sum_{y}\sigma_{ij}(y,x)=\left\langle \sigma_{ij}(y,x)\right\rangle _{V}
\]
fixing root node $x$. Thus $b_{ij}(x)$ is the conditional betweenness
of link $i\rightarrow j$ if the set of shortest paths is restricted
to those terminating at $x$. From this it follows that
\begin{equation}
b_{ij}=\left\langle \left\langle \sigma_{ij}(x,y)\right\rangle _{V}\right\rangle _{V}\label{eq:mueller}
\end{equation}
Comparing~(\ref{eq:mueller}) with~(\ref{eq:marmelade}) we see
that the difference of salience and betweenness is equivalent to the
difference in the shortest path trees $T_{ij}(x)$ and the conditional
betweenness $b_{ij}(x).$ Whereas all links in the shortest path tree
are weighted equally, links with non-zero conditional betweenness
tend to become less central as the links become further separated
from the root node $x$. Formally we can write
\begin{eqnarray}
s_{ij} & = & \left\langle \Theta\left[\left\langle \sigma_{ij}(x,y)\right\rangle _{V}\right]\right\rangle _{V}\nonumber \\
b_{ij} & = & \left\langle \left\langle \sigma_{ij}(x,y)\right\rangle _{V}\right\rangle _{V},\label{eq:yull}
\end{eqnarray}
with $\Theta(x)=1$ if $x>0$ and $\Theta(x)=0$ otherwise.

\subsection{Epidemic simulations}

In order to determine the relevance of link salience to contagion
phenomena on networks, we investigated the correlation of link salience
and the frequency at which links participate in a generic contagion
process that spreads through planar, random triangular networks.

Each network consists of $N=100$ nodes distributed uniformly at random
in a planar disk; the links of the network are given by the Delaunay
triangulation of the nodes. The planar distance between nodes is roughly
proportional to the number of links in a shortest (network) path between
them. A representative example of this type of topology is shown in
Figure~3a. We consider two different weight scenarios:
\begin{enumerate}
\item Quasi-homogeneous weights: Each link is assigned a unit weight $w$
modified by an additive, small perturbation $\xi$
\[
w=1+\xi
\]
where $\xi$ is uniformly distributed in the interval $[-0.01,0.01]$ 
\item Broadly distributed weights: Each link is assigned a random weight
from the distribution with PDF 
\[
p(w)\sim w^{-3}.
\]

\end{enumerate}
We simulate a stochastic Susceptible-Infected (SI) epidemic process.
A single stochastic realization of the process is generated as follows:
Given a network represented by the symmetric weight matrix $w_{ij}$
which quantifies the interaction strength of a pair of nodes, we define
the probability $P_{ij}$ that node $j$ infects node $i$ in a fixed
time interval $\Delta t$ 
\[
P_{ij}=\gamma p_{ij}\quad i\neq j.
\]
where $\gamma\ll1/\Delta t$ is the infection rate, and $p_{ij}=w_{ij}/\sum_{i}w_{ij}$.
Time proceeds in discrete steps; at each step each infected node $j$
chooses an adjacent node to infect at random with probabilities given
by $P_{ij}$. If node $j$ infects a susceptible node $i$, then the
link $(j,i)$ is added to the infection hierarchy $H$, which can
be represented as a matrix $H_{ji}$. In the long time limit every
node is infected, and $H$ is a tree structure recording the first
infection paths from the outbreak location $s$ to every other node.

For a given network, we compute $R=1,000$ different epidemic realizations
with random outbreak locations $s_{k}$, resulting in an ensemble
of infection hierarchies $H_{mn}^{(k)}$. The key question is, how
frequently does a link in the network participate in an epidemic,
and we define the infection frequency of a link as
\[
h_{mn}=\frac{1}{R}\sum_{k=1}^{R}H_{mn}^{(k)}
\]
We compute the infection frequency for 100 random networks under each
weight scenario, and Figure~\ref{fig:Salience-infection} illustrates
the degree to which the directed salience $s_{mn}$ is a predictor
of the dynamic quantity $h_{mn}$. The correlation of $h_{mn}$ with
directed salience and the two measures of centrality we consider here,
weight $w_{mn}$ and betweenness $b_{mn}$, is shown in Table~\ref{tab:salience-infection}.
\begin{table}[H]
\begin{centering}
\begin{tabular}{rcS[table-format=1.3]cS[table-format=1.4]cS[table-format=1.5]} Weight scenario && {$s_d$ vs $h$} && {$b$ vs $h$} && {$w$ vs $h$} \\ \toprule Homogeneous && 0.734 && 0.0756 && 0.00545\\ Broad && 0.803 && 0.329 && 0.393\\ \bottomrule \end{tabular} 
\par\end{centering}

\caption{\label{tab:salience-infection}\textbf{Correlation of other measures
with infection frequency.} The Pearson correlation coefficients of
salience $s_{d}$, betweenness $b$, and weight $w$ with infection
pathway frequency $h$ are shown.}
\end{table}

\bibliographystyle{naturemag}
\bibliography{Salience}

\begin{acknowledgments}
The authors thank O. Woolley-Meza, R. Brune, M. Schnabel, J. Bagrow,
H. Schlämmer and B. Kath for many helpful discussions, and B. Blasius,
A. Motter and B. Uzzi for pointing out and providing some of the data
sets. The authors acknowledge support from the Volkswagen Foundation
and EU-FP7 grant Epiwork.
\end{acknowledgments}

\subsection*{Author contributions}

DG, CT, and DB designed the research. DG and DB developed the theory.
DG, CT and DB analysed the data. DG performed epidemic simulations.
DG and DB wrote the manuscript.

\subsection*{Additional information}

\paragraph*{Competing financial interests:}

The authors declare no competing financial interests.\pagebreak{}
\end{document}